\documentstyle[prl,aps,epsf,multicol,amsfonts]{revtex}
\begin{document}
\draft
\title{Pseudo-spin canting transition in bilayer quantum Hall
  ferromagnets: a self-charging capacitor} 
\author{Leo Radzihovsky}
\address{Physics Department, University of Colorado, Boulder, CO 80309}
\date{Date:\today}
\maketitle

\begin{abstract}
  For sufficiently strong in-plane magnetic field a $\nu_T=1$ bilayer
  quantum Hall pseudo-ferromagnet is expected to exhibit a soliton
  lattice. For sufficiently close layers and large in-plane field, we
  predict this incommensurate ``planar'' phase $P_I$ to undergo a
  reentrant pseudo-spin canting transition to an incommensurate state
  $C_I$, with a finite out-of-plane pseudo-magnetization component,
  corresponding to an interlayer charge imbalance in regions between
  solitons. At $T>0$ the transition is in the 2d compressible Ising
  universality class, and at $T=0$, the quantum transition is in
  heretofore unexplored universality class. The striking experimental
  signatures are the universal nonlinear charge-voltage and in-plane
  field relations, and the divergence of the differential bilayer
  capacitance at the transition, resulting in a bilayer capacitor that
  spontaneously charges itself, even in the absence of an applied
  interlayer voltage.

\end{abstract}

\begin{multicols}{2}
\narrowtext 

There now exists considerable
experimental~\cite{eisenstein92,murphy94,spielman00,spielman01} and
theoretical~\cite{fertig89,macdonald90,wen_zee92,yang_moon94,tunneling01}
evidence for the bilayer phase coherent quantum Hall (QH) state at a
total electron filling fraction $\nu_T=1$, which is driven by Coulomb
exchange interactions and is expected to survive even in the limit of
vanishing interlayer tunneling, $\Delta$.  A number of complementary
pictures of the state, include superfluidity of Chern-Simons composite
boson\cite{wen_zee92}, excitonic superfluid, and
pseudo-ferromagnet\cite{yang_moon94}.

In the latter picture, that we employ here, the $z$ component
$m_z=(n_1-n_2)2\pi\ell^2$ ($\ell = \sqrt{\hbar c/eB}$ is the magnetic
length) of the pseudo-spin magnetization unit vector $\hat{\bf m}={\bf
  m}_\perp + m_z\hat{\bf z}$, is the normalized interlayer imbalance
in electron layer densities $n_{1,2}$, while the azimuthal angle
$\phi=\phi_1-\phi_2$ of ${\bf m}_\perp$ is the difference between the
electron phases $\phi_{1,2}$ in the two layers. Interlayer charging
energy, $2\pi\ell^2\varepsilon_c$, explicitly breaks SU(2) pseudo-spin
symmetry down to $U(1)\times{\mathbb Z}_2$, forcing $\hat{\bf m}$ to
lie in the easy-xy-plane, defined by $m_z=0$. The interlayer tunneling
energy, acts like a pseudo-magnetic field directed along $\hat{\bf
  x}$, further explicitly breaking $U(1)$ symmetry.

\begin{figure}[bth] 
\centering
\setlength{\unitlength}{1mm} 
\begin{picture}(25,80)(0,-12)
\put(-60,-40){\begin{picture}(20,20)(0,0)
\includegraphics{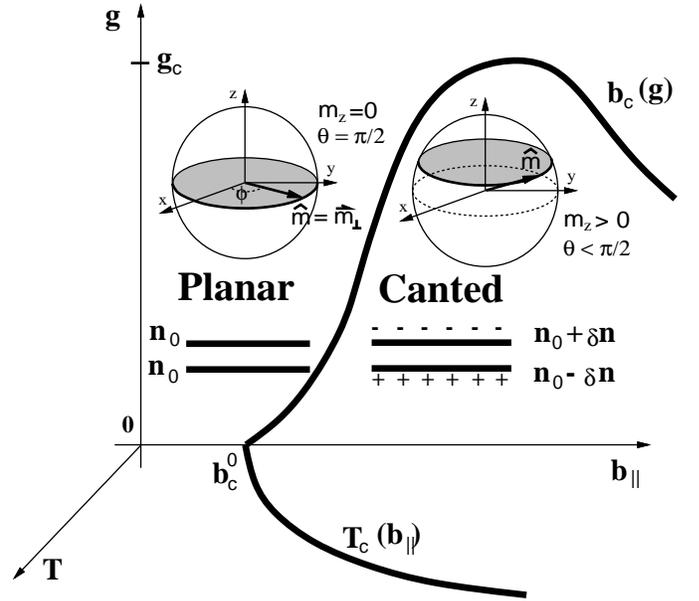}
\end{picture}} 
\end{picture} 
\caption{Phase diagram illustrating a transition between the
  ``planar incommensurate'' $P_I$ and ``canted incommensurate'' $C_I$
  phases of $\nu_T=1$ QH bilayer, with insets showing cartoons of the
  two phases.}
\label{phase_diagram} 
\end{figure}

In this Letter, we predict and explore a striking, reentrant quantum
phase transition from the charge-balanced ``planar'' (P) state into an
interlayer charge-imbalanced ``canted'' (C) state. This novel state is
characterized by the development of $m_z\neq 0$ and corresponds to the
pseudo-magnetization $\hat{\bf m}$ spontaneously canting out of the
easy-xy-plane.  The transition is controlled by dimensionless
parameters $g\equiv 2\pi\ell^2\varepsilon_c/\Delta$ and $b_\parallel\equiv
B_\parallel/B_{CI}$, with a $T=0$ phase boundary $b_c(g)$ illustrated
in Fig.\ref{phase_diagram}, and $B_{CI}$ a critical in-plane field for
the commensurate-incommensurate (CI)
transition.\cite{yang_moon94,CItransition} For a range of parameters,
we find that the transition can be continuous and for $T=0$ is in a
heretofore unexplored $2+1$-d quantum compressible Ising universality
class. At finite $T$ this transition is the 2d compressible Ising
universality class.

Two of many striking and experimentally testable consequences
are (i) the {\em universal nonlinear} behavior of the interlayer
charge imbalance $q(V,B_\parallel)$ with gate voltage $V$ and in-plane
field $B_\parallel$
\begin{equation}
q(V,B_\parallel)=|b_\parallel-b_c|^{\beta}
\tilde{q}(V|b_\parallel-b_c|^{-\chi})\,
\label{q}
\end{equation}
for $b_\parallel>b_c(g)$, leading to a {\em spontaneous} interlayer
charge imbalance $q$, even in the absence of an applied interlayer
voltage, $V\rightarrow0$, i.e., the QH bilayer is a self-charging
capacitor, and
(ii) the concomitant divergence of the interlayer differential
capacitance $C=dq/dV$
\begin{equation}
C(V,B_\parallel)=|b_\parallel-b_c|^{-\gamma}
\tilde{C}(V|b_\parallel-b_c|^{-\chi})\,
\label{C}
\end{equation}
at the quantum ($T=0$) P-C transition, with exponents $\beta=1/2$, $\chi=3/2$,
$\gamma=\chi-\beta=1$, and the dimensionless scaling functions
$\tilde{q}(x)\stackrel{x\rightarrow\infty}{\longrightarrow}
x^{\beta/\chi}$, $\tilde{C}(x)\stackrel{x\rightarrow\infty}
{\longrightarrow}x^{-\gamma/\chi}$ that are universal.  Both
relations should be readily testable in a balanced gate geometry with
the QH bilayer acting as a novel dielectric medium.

We also expect a universal suppression of the quantum Hall gap below
the P-C transition, with $\delta\Delta_{QH}(V,B_\parallel)\propto -
q(V,B_\parallel)^2$, accessible through activated behavior of the
longitudinal resistivity. Similar predictions also hold for the finite
$T$ transition, with $b_\parallel$, $b_c(g)$ replaced by $T$,
$T_c(b_\parallel,g)$, and (for an incompressible soliton lattice) with
{\em exact} critical exponents of the 2d Ising model.

The driving force for the P-C transition is the competition between
the charging (hard axis anisotropy) energy density ${\cal E}_{\rm
charge}=\case{1}{2}\varepsilon_c m_z^2$, minimized by $m_z=0$, and the
exchange energy density ${\cal E}_{\rm
exchange}=\case{1}{2}\rho_s|{\bbox\nabla}\phi|^2$. Because
$\rho_s\approx\rho_s^0 (1-m_z^2)$ (at least in mean-field theory), the
latter can be lowered for any planar twisted state by increasing
$m_z$, via pseudo-spin canting out of the xy-plane, corresponding to
the development of the interlayer charge imbalance.  In the simplest
$\Delta\rightarrow0$ limit, for a uniformly twisted, $\phi=Q x$
staggered current carrying state, this canting instability will
clearly take place for $Q>\sqrt{\varepsilon_c/\rho_s^0}$, corresponding to an
estimate of the ``depairing'' staggered (spin) current ${\bf
J}_s^c={e\over\hbar}\rho_s{\bbox\nabla}\phi={e\over\hbar}\sqrt{\rho_s
\varepsilon_c}\hat{\bf x}$.\cite{kyriakidis01} In contrast, for a finite
tunneling energy $\Delta$ there exist {\em equilibrium} twisted
states, induced by an applied in-plane field ${\bf
B}_\parallel=B_\parallel\hat{\bf y}$.  Although the general driving
mechanism is the same, as we discuss below, the details depend on
whether the canting phase transition takes place out of the planar
commensurate $P_C$ or planar incommensurate $P_I$ (soliton)
state.\cite{yang_moon94}

We begin our analysis by studying the classical energetics of the
quantum Hall bilayer in the presence of an in-plane (physical)
magnetic field ${\bf B}_\parallel=B_\parallel\hat{\bf y}$.  The
appropriate Hamiltonian, in the absense of quenched disorder, and for
a fixed uniform value of $m_z$ is given by
\begin{equation}
\hspace{-0.15in} {\cal H}_{\rm Cl.} = 
{\varepsilon_c\over 2}m_z^2 + {\rho_s\over 2}|\bbox{\nabla}\phi|^2 
- {\Delta\over2\pi\ell^2} \cos\left[\phi\! -\! Q x\right],
\label{Hclassical}
\end{equation}
where $\Delta=\Delta_0(1-m_z^2)^{1/2}$ is the interlayer tunneling
energy,\cite{yang_moon94}\ $Q$ is the in-plane ``magnetic wavevector''
$Q=2 \pi B_{||}d/\phi_0$ associated with $B_\parallel$, and
$\lambda=\sqrt{\rho_s 2\pi\ell^2/\Delta}$ is a screening length for
spatial deformations of $\phi(x)$.  In the {\em commensurate}
state,\cite{yang_moon94,CItransition} stable for
$B_\parallel<B_{CI}=2\phi_0/(\pi^2d\lambda)$, the pseudo-spin
$\hat{\bf m}$ twists {\em uniformly} about $\hat{\bf z}$-axis with
$\phi(x)=Q x$. The corresponding $m_z$ dependent energy density is
given by
\begin{equation}
{\cal E}_C[m_z]= {8\over\pi^2}{\Delta_0\over2\pi\ell^2}
\bigg[b_\parallel^2+(b_0^2-b_\parallel^2)m_z^2
-{\pi^2\over8}(1-m_z^2)^{1/2}\bigg],
\label{E_C}
\end{equation}
where the first term is the planar (i.e., $m_z=0$) exchange energy,
the second consists of the interlayer charging energy, with
$b_0={\pi\over4}\sqrt{2\pi\ell^2\varepsilon_c/\Delta_0}\equiv{\pi\over4}g^{1/2}$
and the reduction of the exchange energy due to canting, and third,
the tunneling energy.\cite{hanna} Simple analysis of ${\cal E}_C[m_z]$
predicts a continuous canting transition from $m_z=0$ planar to
$m_z\neq0$ canted state, at $b_\parallel =
b_c^{(C)}(g)={\pi\over4}\sqrt{g+1}$. Given that it is ${\mathbb Z}_2$
symmetry that is being broken, at $T=0$ ($T\neq0$) we generically
expect such quantum (classical) $P_C-C_C$ transition to be in the
well-studied 3d (2d) Ising universality class, and in the latter case
characterized by {\em exactly} known exponents.  In this scenario,
this $P_C-C_C$ canting transition will be followed by the $C_C-C_I$
transition in a CI universality class, modified by long-range dipolar
interactions\cite{unpublished} and delayed by $m_z\neq0$ to higher
critical in-plane field $B_{CI}=B_{CI}^0(1-m_z^2)^{-1/4}$.

However, because currently studied bilayer
devices\cite{eisenstein92,murphy94,spielman00,spielman01} are
characterized by large $g$, and therefore have $b_c^{(C)}(g)>1$, above
results suggest that the system will first undergo a commensurate
(planar)-incommensurate (planar) ($P_C-P_I$) transition at
$b_\parallel=1$ into a state characterized by a periodic array of 1d
$\hat{\bf y}$-directed solitons.\cite{yang_moon94,CItransition}
Consequently, we need to analyze the $P_I-C_I$ transition from within
such incommensurate soliton state.

Although there has not been any direct experimental evidence, the CI
transition picture was successfully used\cite{yang_moon94} to interpret
a precipitous drop in the QH gap, upon application of sufficiently
strong $B_\parallel$\cite{murphy94}.  In addition to the intrinsic
interest in the $P_I-C_I$ transition, the more easily detected {\em
  canted}-incommensurate ($C_I$) state will also facilitate the
detection and study of the soliton lattice.

The properties of 1d incommensurate soliton state are well documented
in the literature.\cite{CItransition} For $b_\parallel>1$
($Q>Q_{CI}$), the soliton chemical potential is negative and $\hat{\bf
y}$-directed solitons enter the system at a density
$n_s(b_\parallel)$ that continuously increases with $b_\parallel$. In
this state, the phase $\phi_s(x)$ rotates uniformly between solitons,
but slips by $2\pi$ over a soliton width $\lambda$, with ${\bf
m}_\perp(x)$ failing to follow the winding rate $Q$ imposed by
$B_\parallel$. Generically, solitons reduce the energy density
relative to that of the commensurate state with ${\cal E}_I[m_z]=
{\cal E}_C[m_z] - \delta{\cal E}_{\rm solitons}[m_z]$.
%
%

Although, in the absence of fluctuations, exact expressions for
$\phi_s(x)$ and the energy ${\cal E}_I$ (expressible in terms of
elliptic integrals) are available,\cite{CItransition} sufficiently
close to the CI transition thermal fluctuations\cite{comment} {\em
always} qualitatively modify these $T=0$ predictions.  Consequently,
there are at least three regimes above the CI
transition. Asymptotically close to the transition, where the soliton
array is sufficiently dilute, such that $n_s(b_\parallel)<n_T$, with
$n_T$ determined by $n_T\lambda={\rho_s\over
T}e^{-1/(2n_T\lambda)}\approx 1/(2\log{\rho_s/T})$, thermal
fluctuations-induced steric interaction,
$T^2\lambda/(8\rho_s|x_i-x_j|^2)$ dominates over the exponentially
weak $T=0$ interaction ${\rho_s\over\lambda}e^{-|x_i-x_j|/\lambda}$.
Such enhanced soliton repulsion leads to significantly slower (than
$T=0$,
$n_s(b_\parallel)\approx\case{2\pi}{\lambda}|\log{(b_\parallel-1)}|^{-1}$
prediction), power-law increase in soliton density,
$n_s(b_\parallel)\sim{\rho_s\over T \lambda}|b_\parallel-1|^{1/2}$,
and energy density given by
\begin{equation}
\delta{\cal E}_{\rm solitons}^{(T)}[m_z]\sim
{\Delta_0\rho_s^0\over2\pi\ell^2 T}(1-m_z^2)^{15/8}
|b_\parallel-(1-m_z^2)^{-1/4}|^{3/2}.
\label{E_solitonT}
\end{equation}

Ignoring for simplicity a possible intermediate $T=0$ dilute regime,
defined by $n_T\lesssim n_s(b_\parallel)\ll \lambda^{-1}$, for
sufficiently high soliton density $n_T \ll n_s(b_\parallel) \lesssim
\lambda^{-1}$, soliton interaction crosses over to contact
interaction, with energy scale $\rho_s$.  In this dense regime
$n_s(b_\parallel)\sim \lambda^{-1}|b_\parallel-1|$ and soliton energy
density is given by
\begin{equation}
\delta{\cal E}_{\rm solitons}^{(dense)}[m_z]\sim
{\Delta_0\over 2\pi\ell^2}(1-m_z^2)|b_\parallel-(1-m_z^2)^{-1/4}|^2.
\label{E_solitondense}
\end{equation}

Finally in the super-dense limit, $n_s(b_\parallel)\gg \lambda^{-1}$,
$\phi_s(x)$ only shows periodic modulation with vanishing amplitude
and the incommensurate energy density reduces to
\begin{equation}
\delta{\cal E}_{I}^{\rm (super-dense)}[m_z]\approx
{\varepsilon_c\over 2}m_z^2 - {\Delta_0\pi\over 128\ell^2}{1\over b_\parallel^2},
\label{E_superdense}
\end{equation}
with exchange and tunneling energy only leading to a vanishing,
$m_z$-independent correction to the charging energy.

Putting these results together, standard analysis of ${\cal
E}_{I}[m_z]$ shows, that for sufficient small $g$ and sufficiently
large $b_\parallel$ a continuous $P_I-C_I$ transition can take place
at $b_c^{(I)}(g)$, determined by precisely which of the above
regimes it falls into. Because the driving force (the exchange
energy) behind the canting transition vanishes in the
$b_\parallel\gg 1$ limit, Eq.\ref{E_superdense}, we predict that P-C
transition must be reentrant, as illustrated in
Fig.\ref{phase_diagram}.

So far we have restricted our analysis to a spatially uniform
mean-field treatment. While this is appropriate for the commensurate
state and the $P_C-C_C$ transition, this clearly fails in the
incommensurate state, where the soliton array breaks translational
symmetry. A more detailed analysis shows that even within mean-field
theory the canting instability is periodically modulated by
$(\partial_x\phi_s(x))^2$, with canting ($m_z\neq0$) confined to
regions {\em between} solitons, where the twisting of $\phi(x)$ and
therefore the exchange energy cost is highest. One consequence of this
is an upward shift of $b_c^{(I)}(g)$, corresponding to the exchange
energy density cost $n_s\sqrt{\rho_{s z}\varepsilon_c\ell^2}$ associated with
deformation of $m_z(x)$ localized on solitons.

It is noteworthy, that, because $\varepsilon_c$ vanishes while $\rho_s$
saturates in the $d\rightarrow0$ limit\cite{yang_moon94}, in
principle, for sufficiently small interlayer separation $d$, the $P-C$
transition must always take place. Whether it remains continuous,
beyond our above mean-field theory analysis is a more difficult
question. However, estimates of our model parameters from recent
experiments\cite{spielman01}, suggest that currently available bilayer
devices have $g\gg g_c$, and therefore should not display the $P-C$
transition (see Fig.\ref{phase_diagram}).

Transcending these semi-microscopic, model-specific mean-field
considerations, ($\pm m_z$) ${\mathbb Z}_2$ symmetry dictates the form
of the effective classical Hamiltonian
\begin{eqnarray}
H &=& \int d^2r\bigg[{\rho_s^z\over2}|{\bbox\nabla} m_z|^2 +
{\alpha_2\over2}m_z^2+\alpha_4 m_z^4 - {\rm w} m_z^2\partial_x u\nonumber\\
&+&{c_x\over2}(\partial_x u)^2 + {c_y\over2}(\partial_y u)^2\bigg],
\label{H_ClI}
\end{eqnarray}
valid near the finite temperature $P_I-C_C$ transition and on scales
longer than the soliton lattice spacing $n_s^{-1}$. Given our
discussion above, we expect the reduced temperature $\alpha_2\approx
T/T_c(b_\parallel,g) - 1$, the compressional elastic constant $c_x$ to
vanish as $b_\parallel\rightarrow 1^+$ (with precise form depending on
the range of $b_\parallel$), the tilt modulus $c_y\approx\rho_s
n_s/\lambda$, with both approaching $\rho_s$ in the dense regime, and
${\rm w}$ a (pseudo)magneto-elastic coupling of the soliton lattice
phonon $u$ degree of freedom to the local electric dipole moment
$m_z$; because a decrease in the soliton density ($\partial_x u > 0$)
increases the exchange energy of the twisted incommensurate state, we
expect ${\rm w}>0$.

Standard analysis of $H$ in Eq.\ref{H_ClI} predicts that both
$\alpha_4$ and ${\rm w}$ are relevant couplings for $d<4$ and will
therefore lead to non-meanfield critical behavior sufficiently close
to the $P_I-C_I$ transition. Preliminary renormalization group (RG)
analysis in $d=4-\epsilon$ dimensions suggests that the
magneto-elastic coupling ${\rm w}$ can either drive the Ising
transition first order or qualitatively modify it into a new {\em
scalar compressible} Ising model universality
class.\cite{unpublished,BergmanHalperin} If latter scenario survives
down to $d=2$, critical behavior of the magnetization $m_z$ and the
associated magnetic susceptibility leads to the predictions of
Eqs.\ref{q},\ref{C} for the corresponding interlayer charge imbalance
$q$ and the differential capacitance $C$, with gate voltage $V$
playing the role of the associated $\hat{\bf z}$-directed ``magnetic
field''.

We now turn to dynamics. Adapting the imaginary time action of the
pseudo-ferromagnet\cite{yang_moon94} to the incommensurate soliton
state, we find
\begin{eqnarray}
S &=& \int_0^{\hbar/T} d\tau d^2r\bigg[i\gamma m_z\partial_\tau u 
+ {\cal H}[u,m_z]\bigg],
\label{S}
\end{eqnarray}
where the dynamics originates from the Berry's phase ``$p\dot{q}$''
term, that encodes the ferromagnetic precessional dynamics,
$\gamma=\hbar n_s/2\ell^2$, and $\alpha_2 \propto
b_c^{(I)}(g)-b_\parallel$ at $T=0$.

The dynamics of interlayer charge and soliton lattice fluctuations
contained in $S$, Eq.\ref{S} can be probed through a linear response
of interlayer charge imbalance $e\delta n({\bf k},\omega)$ to a
time-dependent interlayer voltage $V({\bf k},\omega)$, applied in a
balanced (i.e., keeping $\nu_T=1$) capacitive geometry.  The relevant
response function is the dynamic dielectric constant $\epsilon({\bf
k},\omega)=d\left({e\over4\pi\ell^2}\right)^2\chi_{zz}({\bf
k},\omega)$, related to interlayer capacitance via a standard relation
$C({\bf k},\omega)=\epsilon({\bf k},\omega)A/d$, both expressible in
terms of the pseudo-spin linear susceptibility, $\chi_{zz}({\bf
k},\omega)=-i\int dt d^2{\bf r} e^{-i\omega t - i{\bf k}\cdot{\bf
r}}\langle[\hat{m}_z({\bf r},t),\hat{m}_z({\bf 0},0)]\rangle_0$.

It is straightforward to compute the $\epsilon({\bf k},\omega)$ away
from the $P_I-C_I$ transition. For the $P_I$ phase we find
\begin{mathletters}
\begin{eqnarray}
\epsilon_P({\bf k},\omega) &=& d\left({e
\over4\pi\ell^2\gamma}\right)^2 {c_x k_x^2 + c_y k_y^2\over -\omega^2
+ \omega_P({\bf k})^2},\\ 
\omega_P({\bf k}) &=& \gamma^{-1}(c_x k_x^2
+ c_y k_y^2)^{1/2}(\rho_s^z k^2 + \alpha_2)^{1/2},
\label{epsilon_P}
\end{eqnarray}
\end{mathletters}
a result resembling that for the $\Delta=B_\parallel=0$
bilayer\cite{yang_moon94}, here, with the translational Goldstone mode
$u({\bf r},t)$ playing the role analogous to the staggered $U(1)$
charge Goldstone phason mode $\phi({\bf r},t)$.  It is noteworthy that
similar to recent theoretical predictions\cite{tunneling01} and
experimental findings\cite{spielman00,spielman01} for interlayer
tunneling, at long wavelengths the peak in the dielectric response
$\epsilon({\bf k},\omega)$ traces out the soliton lattice phonon
dispersion $\omega_P({\bf k})$.

Inside the $C_I$ phase we instead find
\begin{mathletters}
\begin{eqnarray}
\epsilon_C({\bf k},\omega) &=& d\left({e\over4\pi\ell^2\gamma}\right)^2
{c_x k_x^2 + c_y k_y^2\over -(\omega-{{\rm
w}\over\gamma}(\case{|\alpha_2|}{\alpha_4})^{1\over2}k_x)^2 + \omega_C({\bf
k})^2},\nonumber\\ 
&&\\ 
\omega_C({\bf k}) &=& \gamma^{-1}(c_x k_x^2 +
c_y k_y^2)^{1/2}(\rho_s^z k^2 + 2|\alpha_2|)^{1/2},
\label{epsilon_C}
\end{eqnarray}
\end{mathletters}
with $\epsilon({\bf k}\rightarrow0,\omega\rightarrow0)$ and therefore
the associated capacitance (cf. Eq.\ref{C}) diverging as the $P_I-C_I$
transition is approached from above or below.

Near the $P_I-C_I$ transition,
the divergent correlation length
$\lambda_{m_z}=\sqrt{\rho_s^z/\alpha_2}$ leads to breakdown of
perturbation theory, and a full RG analysis of this quantum
compressible Ising transition is in principle necessary. On general
grounds, sufficiently close to $b_c(g)$, we expect $\epsilon({\bf
  k},\omega)$ to display critical scaling
\begin{equation}
\epsilon_{\rm cr.}({\bf k},\omega)=
k_y^{-2+\eta}\,\hat{\epsilon}(k_x/k_y^{z_x},\omega/k_y^{z_\omega},
k_y\alpha_2^{-\nu}),
\label{epsilon_critical}
\end{equation}
where $\nu$ and $z_{x,\omega}$ are the correlation length and
anisotropy exponents, respectively. Preliminary RG
analysis\cite{unpublished} indicates that the upper critical
dimension, below which the quartic ($\alpha_4$) and magneto-elastic
(${\rm w}$) nonlinearities become qualitatively important is
$d_{uc}=2$. Hence, we expect mean-field description with $z_x=1$,
$z_\omega=2$, $\eta=0$, and $\nu=1/2$ to accurately (up to logarithmic
corrections) describe the $P_I-C_I$ transition in these 2d devices.

Up to now we have ignored long-range part of the electrostatic dipolar
interaction $\int_k U(k)|m_z(k)|^2$, with kernel
$U(k)=\case{\pi\varepsilon_d}{3d^2}(1-e^{-k d})/k$, and
$\varepsilon_d=\left({e
    d\over4\pi\ell^2}\right)^2{1\over4\pi\epsilon}$ is the dipolar
energy per unit of length. As in real bulk magnets this interaction
clearly favors a development of anti-aligned dipolar domains at the
shortest possible length scales and therefore competes with the
exchange energy $\case{1}{2}\rho_s^z|{\bbox\nabla}m_z|^2$, minimized
by spatially homogeneous $m_z({\bf r})$. In contrast to 3d systems,
where domain size scales as $\sqrt{L}$, it is easy to show that in our
2d geometry, sufficiently below the $P_I-C_I$ transition, the domain
length is $\xi_d\approx d e^{\varepsilon_{\rm exch.}/\varepsilon_d}$,
($\varepsilon_{\rm exch.}\approx\rho_s^z/\lambda_{m_z}$) and therefore
is a length that can in principle be tuned from bilayer thickness to a
macroscopic length. Sufficiently close to the transition, the
long-range dipolar interaction will always become important, and will
lead to the canting instability in $m_z({\bf r})$ to take place at a
finite wavevector $k_d$, determined by ${\rho_s^z\over\varepsilon_d d}
= {\pi\over 3 (k_d d)^3}\left[1-(1+k_d d)e^{-k_d d}\right]$. Although
the mean-field predictions of the thermodynamics at the transition
should remain unchanged, the {\em uniform} ($k_x=0$) pseudo-spin
susceptibility, determining the capacitance $C$ will no longer
diverge. For sufficiently large domains $\xi_d$, we expect
fluctuations enhancement of $C$ near the transition, but with its
divergence now cutoff by the dipolar domain size, and therefore with
its peak scaling according to $k_d^{-(2-\eta)/z_x}$.

We have argued that in bilayer QH ferromagnets, for
sufficiently strong field and close interlayer separation a quantum
interlayer charging transition must take place. 
While arguments in favor of the
transition and its striking properties 
are quite general, it is unlikely that our
mean-field estimates of the phase boundary are quantitatively
trustworthy. Microscopic calculations, exact diagonalization
and quantum Monte Carlo studies are necessary to accurately
determine the details of the phase diagram proposed here. We 
hope that the signatures of the $P-C$ transition
studied here will stimulate experimental efforts to develop bilayer
devices, where it can be observed.

The author thanks L. Balents, A. Dorsey, S. Girvin, 
and J. Toner for discussions, and acknowledges support by the 
NSF DMR-9625111, and by the Sloan and Packard Foundations.

\end{multicols}
\end{document}